\documentclass[11pt,preprint]{aastex}
\slugcomment{Accepted AJ, 2009 August 15}
\shorttitle{MIPS Observations of S-COSMOS}
\shortauthors{Frayer et al.}
\begin{document} 

\title{{\em Spitzer} 70 and 160~$\mu$\lowercase{m}
Observations of the COSMOS Field}

\author{D.\ T.\ Frayer\altaffilmark{1}, 
D.\ B.\ Sanders\altaffilmark{2},
J.\ A.\ Surace\altaffilmark{3},
H.\ Aussel\altaffilmark{4}, 
M.\ Salvato\altaffilmark{5},
E.\ Le Floc'h\altaffilmark{2},
M.\ T.\ Huynh\altaffilmark{1},
N.\ Z.\ Scoville\altaffilmark{5},
A.\ Afonso-Luis\altaffilmark{6},
B.\ Bhattacharya\altaffilmark{1},
P.\ Capak\altaffilmark{3}, 
D.\ Fadda\altaffilmark{1}, 
H.\ Fu\altaffilmark{5}, 
G.\ Helou\altaffilmark{1},
O.\ Ilbert\altaffilmark{2},
J.\ S.\ Kartaltepe\altaffilmark{2},
A.\ M.\ Koekemoer\altaffilmark{7},
N.\ Lee\altaffilmark{2},
E.\ Murphy\altaffilmark{3},
M.\ T.\ Sargent\altaffilmark{8},
E.\ Schinnerer\altaffilmark{8},
K.\ Sheth\altaffilmark{3},
P.\ L.\ Shopbell\altaffilmark{5},
D.\ L.\ Shupe\altaffilmark{1},
L.\ Yan\altaffilmark{3}}

\altaffiltext{1}{Infrared Processing and Analysis Center, California
Institute of Technology 100-22, Pasadena, CA 91125, USA}

\altaffiltext{2}{Institute for Astronomy, 2680 Woodlawn Drive,
University of Hawaii, Honolulu, HI 96822, USA}

\altaffiltext{3}{{\em Spitzer} Science Center, California Institute of
Technology 220--06, Pasadena, CA 91125, USA}  

\altaffiltext{4}{CNRS, AIM-Unit\'{e} Mixte de Recherche
  CEA-CNRS-Universit\'{e} Paris VII-UMR 7158, F-91191 Gif-sur-Yvette,
  France.}

\altaffiltext{5}{Astronomy Department, California Institute of Technology
105--24, Pasadena, CA  91125, USA} 

\altaffiltext{6}{Instituto Astrofiscia de Canarias, Via Lactea, 38200
La Laguna, S/C de Tenerife, Spain}

\altaffiltext{7}{Space Telescope Science Institute, 3700 San Martin
Drive, Baltimore, MD 21218, USA}

\altaffiltext{8}{Max-Planck-Institut f\"{u}r Astronomie,
  K\"{o}nigstuhl 17, D-69117 Heidelberg, Germany}

\begin{abstract}

 We present {\em Spitzer} 70 and 160~$\mu$m observations of the COSMOS
 {\em Spitzer} survey (S-COSMOS).  The data processing techniques are
 discussed for the publicly released products consisting of images and
 source catalogs.  We present accurate 70 and 160~$\mu$m source counts
 of the COSMOS field and find reasonable agreement with measurements
 in other fields and with model predictions.  The previously reported
 counts for GOODS-North and the extragalactic First Look Survey are
 updated with the latest calibration, and counts are measured based on
 the large area SWIRE survey to constrain the bright source counts.
 We measure an extragalactic confusion noise level of $\sigma_c =
 9.4\pm3.3$\,mJy ($q=5$) for the MIPS 160~$\mu$m band based on the
 deep S-COSMOS data and report an updated confusion noise level of
 $\sigma_c = 0.35\pm0.15$\,mJy ($q=5$) for the MIPS 70~$\mu$m band.

\end{abstract} 


\section{Introduction}

The Cosmic Evolution Survey (COSMOS) is a deep multi-wavelength
wide-area (2 deg$^2$) program for studying the evolution of galaxies
and active galactic nuclei (AGN) \citep{sco07}.  The COSMOS {\em
Spitzer} (S-COSMOS) survey is comprised of the Infrared Array Camera
(IRAC, 3.6--8$\mu$m) and Multiband Imaging Photometer for {\em
Spitzer} (MIPS, 24, 70, and 160~$\mu$m) data \citep{san07}.  This
paper presents the far-infrared (FIR) 70 and 160~$\mu$m MIPS
observations of the field.  Although the mid-infrared (MIR) 24~$\mu$m
array is more sensitive to the detection of distant galaxies than the
MIPS-Germanium (MIPS-Ge) 70 and 160~$\mu$m detectors, the 24~$\mu$m
data are biased toward warm AGN and are affected by broad mid-infrared
PAH emission and silicate absorption features redshifted into the
band.  The strong MIR spectral features along with the large
variations of the FIR/MIR continuum ratios \citep[e.g.,][]{dal05}
yield highly uncertain bolometric corrections.  The long-wavelength 70
and 160~$\mu$m observations directly measure the FIR peak of the
spectral-energy distributions (SEDs) for redshifts $z\la 1.5$ and are
key for constraining the total infrared luminosities and
star-formation rates of galaxies within the COSMOS field.  The MIPS 70
and 160 $\mu$m data provide an important piece of the puzzle in the
quest of understanding galaxy evolution.

The goal of the data paper is to document the data products to
facilitate the ongoing research of the COSMOS field.  The data
products and a description of the observations and data reduction are
provided as part of the large public repository of multi-wavelength
data for the COSMOS field ({\em Hubble Space Telescope} Advanced
Camera for Surveys (ACS), Koekemoer et al. 2007; radio, Schinnerer et
al. 2007; X-ray, Hasinger et al. 2007; and optical and near-infrared,
Capak et al. 2007).  We present two scientific results here: (1) the
70 and 160~$\mu$m source counts, and (2) the measurement of the
confusion noise for the MIPS 160~$\mu$m band.

\section{Observations}

The MIPS S-COSMOS observations were carried out in 4 campaigns from
2006 January through 2008 January (Table 1).  The project ({\em
Spitzer} programs 20070 and 30143) represents over 450\,hr of MIPS
observations.  All observations were taken using the scan mapping mode
during nominal ``Cold'' MIPS campaigns (telescope temperatures low
enough to yield good quality 160~$\mu$m data).  The initial
observations taken in 2006 Cycle-2 are described in \cite{san07}.  The
Cycle-2 observations were comprised of a shallow wide-area ($1.75\deg
\times 1.97\deg$) survey to quantify the level of cirrus within the
field and a small ($0.5\deg \times 0.33\deg$) deep ``test'' field.
After the successful completion of the Cycle-2 program, deep
observations over the entire COSMOS field were carried out in Cycle-3
(depth of about 2800\,s, 1350\,s, and 270\,s in the MIPS 24, 70, and
160~$\mu$m bands respectively).

The Cycle-3 astronomical observational requests (AORs) were optimized
specifically for the MIPS-Ge bands, without compromising the 24 $\mu$m
data.  In contrast, some early MIPS programs of other groups were
designed for the 24~$\mu$m band at the expense of the MIPS-Ge bands.
The forward and return scan legs were offset by 148$\arcsec$ which
provides sufficient overlap for the 70~$\mu$m array.  Cross-scan
dither offsets of $\pm$0, 42, 83$\arcsec$ and in-scan dither offsets
of $\pm$0, 18$\arcsec$ were used between multiple maps to account for
the unusable parts of the MIPS-Ge arrays and the unobserved central
row of the 160~$\mu$m footprint.  The values of the cross-scan dithers
were also chosen to avoid the overlap of 24~$\mu$m readouts between
consecutive maps.  We carried out both forward and reverse scan maps
to help characterize the long-term transients of the MIPS-Ge
detectors.
 
The majority of the data were taken in slow scan mode.  At the slow
scan speed ($2.6\arcsec {\rm s}^{-1}$), each AOR consisted of 4 scan
legs of 1.5\,deg.  Each AOR mapped 1.5\,deg$\times 592\arcsec$, and 10
AORs were used to map the field once.  In total, 13 slow maps
(29.5\,hr per map) were carried out in Cycle-3, along with one map at
the medium scan rate (12.5\,hr) to complete the awarded time.  Five
AORs were lost due to satellite downlink issues in the second epoch of
Cycle-3 and were re-observed in early 2008 (Cycle-3c, Table~1).

The scan direction of MIPS is determined by the date of observation,
and the observations were carried out on the days that minimized the
zodiacal light.  Since the field is near the ecliptic plane, the
zodiacal background contributes significantly to the total noise
budget for the MIPS 24 and 70~$\mu$m bands.  The zodiacal light is not
significant at 160~$\mu$m.  The galactic cirrus level is low in the
direction of the COSMOS field \citep{san07} and is not the dominant
source of confusion noise within the MIPS-Ge bands.  Figures~1 and 2
show the final MIPS 70 and 160~$\mu$m images, combining all epochs
(Table~1).  Since the 70~$\mu$m and 160~$\mu$m arrays are on the
opposite side of the MIPS field of view, the overscan regions yield
slightly non-symmetric coverage.  For both Cycle-3a and Cycle-3b, the
entire ACS field was observed, and the MIPS over-scan regions provide
coverage for the IRAC data outside of the ACS field.

\section{Data Reduction}

The raw MIPS-Germanium 70 and 160~$\mu$m (MIPS-Ge) data were
downloaded from the {\em Spitzer} Science Center (SSC) archive and
were reduced using the Germanium Reprocessing Tools (GeRT, version
20060415) and additional specialized scripts developed for processing
the MIPS-Ge survey data.  The GeRT uses an offline version of the SSC
pipeline to produce the basic calibrated data products (BCDs).  The
basic MIPS-Ge processing steps are discussed by \cite{gor05} and
within the GeRT documentation.  The processing for S-COSMOS made use
of lessons learned from the processing of the extragalactic First Look
Survey \citep[xFLS,][]{fra06a} and the deep GOODS-North observations
\citep{fra06b}.  Additional processing enhancements were derived here
using the S-COSMOS data.

The main processing steps were carried out in the following order: (1)
calculation of the data ramp slope, (2) stimulator-flash
interpolation, (3) improved stimulator-flash response solution, (4)
calibration of the slope image to yield the BCD product, (5) enhanced
filtering of the BCD product, (6) data co-addition, (7) identification
of bright sources, (8) masking bright sources and re-calculation of
the filtering corrections, (9) final data co-addition, and (10) final
source extraction. For comparison, the on-line SSC pipeline only
performs steps (1), (2), (4), basic filtering, and step (6), and the
GeRT includes software for steps (1), (2), (4), and (8).  The 10
processing steps are summarized in the following sub-sections.

\subsection{Basic Processing}

The optimal processing for step (1) depends on the background, the
length of the data ramp (MIPS-Ge data are recorded with
non-destructive reads sampled at 0.131~s), and the rate of cosmic rays
at the time of the observations.  The SSC on-line pipeline is tuned
for the short data ramps (3\,s and 4\,s).  We tuned the pipeline
modules (cosmic ray detection and removal and slope estimation) of
step (1) to minimize the noise level for the longer 10\,s ramps of the
S-COSMOS data.  The tuning solutions for 70~$\mu$m are similar to
those derived for the GOODS-North photometry data which have the same
ramp length \citep{fra06b}.  MIPS-Ge uses stimulator-flash
observations to track the response of each detector as a function of
time.  After the calculation of the initial stimulator-flash solution,
which is basically a linear interpolation between stimulator-flash
measurements (step 2), we removed outlier values and re-derived a
smoothed stimulator-flash solution (step 3).  For 70~$\mu$m, the
stimulator-flash response function was smoothed by about 2 minutes
(slightly longer than the stimulator-flash cycle) to provide the
lowest noise.  For 160~$\mu$m, the solution was smoothed by about 8
minutes to yield the best results.

In step 4, the BCD data are calibrated as $BCD(t)=FC[U(t)/SR(t) -
DARK]/IC$, where $U(t)$ is the uncalibrated slope image, $SR(t)$ is
the stimulator-flash response solution derived in step (3), and $DARK$
is the dark calibration file. The $IC$ calibration file is the
illumination correction which corrects for the flat-field response
and the non-uniformity of the stimulator flash \citep{gor05}.  The
flux conversion factor ($FC$) converts the instrument units into
physical surface brightness units of MJy\,sr$^{-1}$.  For
self-consistency, we adopt the same $DARK$ and $IC$ files used for the
official calibration of the MIPS instrument \citep{eng07, gor07,
sta07}.  The calibration $FC$ values of 702 MJy\,sr$^{-1}$ per
MIPS-70-unit \citep{gor07} and 41.7 MJy\,sr$^{-1}$ per MIPS-160-unit
\citep{sta07} are adopted.  The calibration of MIPS is based on
stellar SEDs ($S_{\nu} \propto \nu^{2}$).  We have placed the data on
a constant $\nu\,S_{\nu}$ scale by dividing the data by the color
correction factors of 0.918 and 0.959 for the 70 and 160 $\mu$m bands,
respectively \citep{sta07}.  These color corrections are appropriate
(accurate to better than 2\%) for a wide range of galaxy and AGN SEDs
($S_{\nu} \propto \nu^{-\alpha}$, $\alpha =0$--3) across the filter
bandpass.

\subsection{Data Filtering}

Optimization of the processing steps (1-4) provided sensitivity
improvements of about 20\% in comparison to the default parameters,
while the filtering (step 5) can provide more than a factor of two
improvement in point-source sensitivity.  Filtering is key in the
removal of systematic instrumental effects which impede the ability to
integrate down with deep observations.  The two main artifacts
impacting MIPS-Ge data are the latents due to the stimulator flashes
and variations of the slow response ($> 2$\,min) as a function of
time.  The slow response is removed at 70 and 160~$\mu$m by
subtracting a running median per detector as a function of time, i.e.,
a high-pass filter.  The latent artifacts due to the stimulator
flashes are not fully removed by a simple high-pass filter, since
these variations occur on shorter time-scales.  At 70 $\mu$m the
stimulator-flash artifacts are correlated by column.  Since the scan
direction is nearly along the columns of the array, these artifacts
contribute to the streaking within the maps if not corrected.  We
remove the column residuals by subtracting the median of the values
along each column for every BCD at 70~$\mu$m.  The combination of the
column median filter and a high-pass median time filter per detector
removes the instrumental artifacts at 70~$\mu$m.

There is no equivalent column filter at 160~$\mu$m to remove the
high-frequency (short-time scale) latent images introduced by the
stimulator-flashes.  Fortunately at 160~$\mu$m, these artifacts are
repeatable and can be determined by stacking the data as a function of
DCENUM (Data-Collection-Event Number) within the stimulator-flash
cycle.  Since the scan-mirror position also varies with DCENUM within
the stimulator-flash cycle, we stacked the data per AOR for each
scan-mirror position and took the median value of the stack to derive
the correction as a function of detector and scan-mirror position.
These corrections were subtracted from the BCDs to remove artifacts
due to the stimulator-flashes.  The combination of this stacking
correction and a high-pass median time filter removes the instrumental
artifacts at 160~$\mu$m.

With the S-COSMOS data, we tuned the filtering techniques (step 5) to
minimize the noise in the output maps.  For both 70 and 160~$\mu$m,
the short-time scale stimulator-flash artifacts were removed first,
followed by a high-pass median time filter to remove the longer
time-scale transients.  We adopted a time filter width of 12 frames (2
minutes) to yield the best results.  After the initial filtering, the
brightest sources have negative ``side-lobes'' in the maps since
bright sources bias the calculation of the median for neighboring
pixels.  To remove these filtering artifacts, the filtering was done
in two passes.  The data from the first filtering pass (step 5) were
co-added (step 6), and sources were extracted (step 7) to find the
location of the bright sources.  The source positions within the
original BCDs were masked and new filtering corrections were
calculated in a second pass, ignoring the pixels containing sources.
This two-pass filtering technique minimizes the artifacts while
preserving the point-source calibration.  After the second filtering
pass (step 8), the data were co-added to produce the final maps (step
9, Sec.~3.3) and sources were extracted to produce the catalogs (step
10, Sec. 3.4).

\subsection{Imaging}

The SSC mosaicking software \citep[MOPEX, Version 16.3.7,][]{mak05} was used
to combine the data and make the final images.  A fast plane-to-plane
coordinate transformation method was used to project the data onto the
sky \citep{mak04} with the default MOPEX interpolation scheme.  We
carried out the imaging steps following the techniques discussed for
the xFLS MIPS-Ge data \citep{fra06a}.  An important improvement
available after the processing of the xFLS data is a more robust
outlier rejection technique within MOPEX.  The updated method rejects
data around the median of a data stack for each sky pixel instead of
rejecting data with respect to the average of the data stack.  This
enables more aggressive outlier rejection without compromising the
calibration of point sources.  The best sensitivity was obtained using
rejection thresholds of $\pm 2.5 \sigma$.  The new outlier rejection
method improved the sensitivity in the maps by about 5\%.

The pipeline uncertainties of the BCDs and integration times were not
used as weights in the co-addition of the data; all of the data
(155,411 BCDs in total) flagged as good were given equal weight.  Bad
data flagged on a detector basis during the pipeline processing or
identified as outliers by MOPEX were not included.  The vast majority
of the data were taken with 10\,s integrations and no correction is
needed for the 4\,s BCDs (all BCDs are calibrated correctly in
MJy\,s$^{-1}$).  The pipeline uncertainties (calculated from the
formal error propagation of the pipeline steps) do not fully represent
the actual noise characteristics of the data, and underestimate the
real noise slightly in low background regions.  The main utility of
the pipeline uncertainties for these data is to provide a lower limit
to the input noise for the MOPEX outlier rejection algorithm.

\subsection{Source Detection and Extraction}

Sources were detected and extracted from the final images using the
Astronomical Point-Source Extraction (APEX) tools within the MOPEX
software package and using additional specialized scripts.  For
optimal source detection and extraction, accurate background
subtraction and noise estimates are needed.  The filtering process
(steps 5\&8) yields a small systematic negative offset of the
background level in the image.  This offset was estimated by taking
the median of the image within the central regions after masking
sources detected at levels greater than $3\sigma$.  After the removal
of the global offset level ($-0.05$~MJy\,sr$^{-1}$ at 70 $\mu$m and
$-0.07$~MJy\,sr$^{-1}$ at 160 $\mu$m), there are still local
background fluctuations across the image depending on the local
density of sources (both from detected sources and
un-detected/confused sources associated with infrared galaxies seen at
24 $\mu$m).  The local background level was derived by taking a median
within a box around each pixel (after masking sources detected at
levels greater than $3\sigma$).  For source detection, we used a small
box with a linear size of 5 Full-Width Half-Maximum (FWHM) widths to
remove the local background, and for source extraction (fitting) we
used a larger background box with a size of 9--10 FWHM widths to
conserve the calibration.

Several types of noise images can be produced by the MOPEX and APEX
software, but none are optimal for these data.  The ``std'' noise
image produced by MOPEX represents the empirical scatter from the
repeated observations per sky pixel divided by the square-root of the
number of good observations.  For deep observations the ``std'' file
underestimates the true noise since it does not account for the
pixel-to-pixel correlated noise or the confusion noise.  APEX computes
the spatial pixel-to-pixel noise (``noise'' file) by calculating the
noise within a local box surrounding each pixel after rejecting
positive outliers.  The output ``noise'' image has variations that
depend on the outlier parameter and the number of sources within the
local box.  To avoid local biases due to sources, we derived a
``noise'' image after the extraction of sources, adopting a box size
with a linear scale of 9-10 FWHM widths (the same size used for the
local background subtraction for source fitting).  To preserve both
the small and large scale spatial variations of the uncertainty across
the science image, both the ``std'' (representing small-scale
variations) and ``noise'' (representing large-scale variations) files
were used.  The quality of ``std'' and ``noise'' images was first
improved by smoothing the images by 1 and 3 FWHM widths, respectively,
and then combined in quadrature.  Equal weights were given to the
``std'' and ``noise'' images at 70~$\mu$m.  Since the confusion noise
is important at 160~$\mu$m, the weights at 160~$\mu$m were based on
the relative contributions of the instrument and confusion noise
(Sec. 5.2).  The std image was given a weight corresponding to the
instrument noise, and the noise image was given a weight corresponding
to the confusion noise to produce the combined uncertainty file at
160~$\mu$m.  The median level of the combined uncertainty file was
scaled to match the total average noise level derived from fitting the
$1\sigma$ width of a Gaussian to the data histogram of the image after
source extraction.  Source extraction and noise estimates were
repeated until the results converged.

After proper background subtraction and deriving an accurate
uncertainty image, sources were detected using the APEX peak
algorithm.  Peaks with a signal-to-noise ratio (SNR) of greater than 3
were fitted using the Point-source Response Function (PRF, which
includes the effects of the detector size and the adopted sub-sampling
of the detectors) image.  At the spatial resolution of the MIPS-Ge
bands, the PRF is stable.  At 70 $\mu$m we adopt the PRF (FWHM
$=18.6\arcsec$) made previously from the xFLS data \citep{fra06a}.  At
160~$\mu$m we made a new PRF using the COSMOS, xFLS, and the data from
all of the fields from the Spitzer Wide-area Infrared Extragalactic
Legacy Survey \citep[SWIRE,][]{lon04}.  None of the individual fields
has a large number of isolated 160~$\mu$m sources with high SNR to
make a high-quality PRF.  At 160~$\mu$m, the brightest sources ($\ga
2$ Jy) cannot be used due to the non-linearity of the detectors
\citep{sta07}.  In total, we used 149 isolated sources from the
Spitzer 160~$\mu$m surveys which have SNR $\ga30$ and $S160 <
1$\,Jy\footnote{Throughout this paper the flux densities of the MIPS
bands are defined as $S160=S_{\nu}(155.9\mu$m),
$S70=S_{\nu}(71.4\mu$m), and $S24=S_{\nu}(23.7\mu$m).}  to produce an
empirical PRF (FWHM $=39\arcsec$).

The detection and extraction of sources was done in multiple passes.
Initial source lists were visually analyzed and cleaned to remove
sources with low coverage and potential spurious sources in the Airy
rings around bright sources. The detection table was updated to remove
bad sources, and source fitting was redone.  In the few cases where
the detection routine failed in the proper de-blending of two or more
sources, the detection table was modified, and source fitting
measurements were redone using better initial source positions.

\subsection{Astrometry}

The absolute pointing reconstruction of the {\em Spitzer} telescope is
typically better than 1$\arcsec$.  The pointing uncertainties are much
less than the large FWHM width of the data ($18.6\arcsec$ at 70~$\mu$m
and $39\arcsec$ at 160~$\mu$m).  We verified the pointing solutions by
comparing the positions of the 70~$\mu$m sources with their VLA radio
counterparts \citep{sch07}.  For the approximately 400 sources
detected at SNR$>5$ at both 1.4~GHz and 70~$\mu$m, we find an average
positional difference ({\em Spitzer} - VLA) of $\Delta{\rm
RA}=-0.08\pm0.08\arcsec$ and $\Delta{\rm Dec}=0.17\pm0.08\arcsec$.
These small offsets are well within the scatter measured for
individual sources of 1--2$\arcsec$, so no positional corrections were
made to the data.

\subsection{Calibration}

MIPS is calibrated assuming the point-spread functions (PSF, which
does not include the effect of the detector size) calculated using the
Spitzer TinyTim (S-TinyTim) models (Krist 2002).  The modeled PSFs
have been shown to match the observations for all bands \citep{eng07,
gor07, sta07}.  An important aspect of the MIPS PSFs is that
significant flux arises outside of the first Airy ring on large
spatial scales.  We corrected the derived flux densities for emission
outside of the adopted PRF images using the S-TinyTim models.  For the
adopted PRFs ($87\arcsec \times 87\arcsec$ at 70~$\mu$m and
$190\arcsec \times 190\arcsec$ at 160~$\mu$m), we derive an aperture
correction of 1.15 at both wavelengths based on modeled PSFs assuming
a constant $\nu\,S_{\nu}$ SED.  Empirically, it is difficult to
measure the level of emission outside of the first Airy ring for
MIPS-Ge data.  However, the calibration factors assume the modeled
PSFs out to very large spatial scales, and we must apply this
correction for consistency.  We verified the calibration consistency
of our techniques using archived calibration observations taken over
the lifetime of the mission.  For comparison, observations of the
calibration star HD180711 \citep[$S70=447.4$\,mJy,][]{gor07} were used
at 70~$\mu$m, while observations of the ULIRG IRAS 03538--6432 were
used at 160~$\mu$m \citep[($S160 = 1.04$\,Jy)]{sta07,kla01}.  Using an
aperture correction for a modeled PSF with a stellar SED and similar
reduction techniques carried out for S-COSMOS, we measure a flux
density of $450\pm18$\,mJy for HD180711.  This is consistent with
expectations (flux density ratio of $1.01\pm0.04$).  At 160~$\mu$m we
measure a flux density ratio of $0.97\pm0.05$ compared to expectations
for IRAS 03538--6432 using similar reduction techniques carried out
for S-COSMOS.  These results suggest that the calibration of the
S-COSMOS data is consistent with the MIPS calibration papers; the
absolute calibration uncertainty of MIPS is 5\% at 70~$\mu$m
\citep{gor07} and 12\% at 160 $\mu$m \citep{sta07}.

\subsection{Corrections for Completeness and Eddington Bias}

The completeness levels were estimated by simulations and calculations
based on the SNR threshold.  For the simulations, sources were
injected at random locations into the images and extracted using the
same techniques adopted for the production of the catalogs.  For high
SNR $\geq 5.0$, the SNR threshold itself is the dominant effect in the
determination of the completeness level.  At lower SNR, other effects
such as the details associated with source detection and fitting
become significant.  The completeness levels vary significantly across
the image as a function of coverage and flux density.  Based on the
SNR $\geq 5.0$ threshold, the average effective completeness level
across the image can be calculated as a function of flux density as
the fractional area within the image for which $S+I \geq 5.0 U$, where
$S$ is the flux density, $I$ is the science image, and $U$ is the
uncertainty image.  Figures~3\&4 show completeness calculations as a
function of flux density and coverage.  The simulations match the
expected curves very well at 70~$\mu$m for both the deep and typical
regions within the image.  For the deep region at 160~$\mu$m, the
simulated completeness values are relatively noisy and are lower than
expected for 60--90\,mJy, potentially due to the effects of confusion
and/or the small number of independent beams within the simulated
area.  The completeness calculations at 160 $\mu$m for the nominal
coverage of greater than 20 are similar to the simulations and the
calculations for the typical range of coverage values (25--33).  For
the derivation of the S-COSMOS source counts (Sec.~5.1), we adopt the
completeness curves for coverages greater than the nominal values of
100 and 20 at 70 and 160~$\mu$m, respectively (solid lines,
Fig. 3\&4).

In addition to the completeness corrections, the counts are affected
by the Eddington bias (flux boosting).  At faint flux densities the
observed flux densities are slightly higher on average than the true
flux densities since sources on positive noise features are
preferentially selected.  This effect of "flux boosting" is not as
important for SNR $\geq5$ as for lower SNR, but would still yield a
small systematic bias in the measured counts if not corrected.  The
same simulations used to help quantify the completeness corrections
were used to check the importance of the Eddington bias.  The ratio of
the observed to input flux densities ($S_{\rm obs}/S_{\rm true}$) of
each bin were measured for both the deep region and the wide area with
"full-blown" simulations that injected sources at random positions
within the image (one at a time) and exacted the output flux densities
using the same methods adopted to produce the source catalogs.  Given
the variation of coverage in the data, it is difficult to obtain
sufficient statistics as a function of flux density and coverage
across the image using these full-blown simulations (as was the case
for the completeness estimates).  Instead, we carried out simpler
calculations that are consistent with the full-blown simulations, but
are significantly more accurate.  Input flux densities ($S_{\rm
true}$) with the same power-law distribution as the observed source
counts were randomly added to the noise distribution given by the
uncertainty image, and the output observed flux densities were
derived.  The effective Eddington bias of $S_{\rm obs}/S_{\rm true}$
was calculated over the entire image for each flux density bin for the
adopted SNR $\geq 5$ cut and used to correct the observed flux
densities for the derivation of the counts (Sec.~5.1).  This method
fully accounts for the variation of coverage and noise across the
image.

\section{Products}

The S-COSMOS 70 and 160~$\mu$m products are available on-line at the
NASA Infrared Science Archive (IRSA) at the Infrared Processing and
Analysis Center (IPAC).  The products described here are version~3 of
the data.  Version~1 and version~2 were early quick-look products
based on simplified reductions of sub-sets of the observations.
Version~3 represents the combination of all data from the S-COSMOS
MIPS program and is the first version to be calibrated properly and
processed with the best known data reduction techniques.

\subsection{Images}

The science images have been background subtracted with the removal of
a global offset, but the local background fluctuations have not been
removed (Sec. 3.4).  The images are in surface-brightness units of
MJy\,sr$^{-1}$ assuming the latest calibration and have been
color-corrected to match SEDs with a constant $\nu\,S_{\nu}$ scale
(Sec. 3.1).  The uncertainty images have also been color-corrected and
are in units of MJy\,sr$^{-1}$.  The uncertainty images ($1\sigma$)
represent both the small-scale and large-scale spatial noise
properties associated with the science image (Sec. 3.4).

The coverage maps provide the effective number of observations (after
data editing) per point on the sky for the science images.  At 70
$\mu$m for the nominal coverage of greater than 100, the median
effective exposure time is 1350\,s while the ultra-deep region with a
coverage of greater than 250 has an exposure time of 2800\,s.  At 160
$\mu$m for the nominal coverage of greater than 20, the median
effective exposure time is 273\,s, while the ultra-deep region with a
coverage of greater than 50 has an effective exposure time of 567\,s.

Table~2 lists the image properties and sensitivities.  The
sensitivities represent the total noise, including confusion.  The
surface-brightness noise for the adopted pixel scale was derived from
fitting the $1\sigma$ width of a Gaussian to the data histogram of the
image after source extraction.  To derive the effective point-source
noise, including the effects of correlated noise between the pixels,
we carried out aperture measurements at random locations within the
residual image after the extraction of sources.  We derive conversion
factors between the point-source noise and surface-brightness noise of
$13.3\pm1.3$\,mJy\,(MJy\,sr$^{-1})^{-1}$ and
$74\pm7$\,mJy\,(MJy\,sr$^{-1})^{-1}$ for the 70~$\mu$m and 160~$\mu$m
images respectively.  The average point-source noise ($1\sigma$) is
1.7\,mJy and 13\,mJy at 70 and 160~$\mu$m, respectively.  These
point-source values include the aperture correction of 1.15 for
emission outside of the measured PRF (Sec. 3.6).

\subsection{Catalogs}

Point-source catalogs were made for a SNR $\geq 5.0$.  In total, 1512
sources at 70~$\mu$m (Table~3) and 499 sources at 160~$\mu$m (Table~4)
are cataloged.  The catalogs are single-band source lists and are
independent from each other and the MIPS 24 $\mu$m data.  Although the
catalogs are not biased by data at other wavelengths, we did use the
24 $\mu$m and radio data to confirm the reliability of the catalogs.
Within the central area cataloged at 24 $\mu$m, only seven 70~$\mu$m
sources do not have a 24 $\mu$m counterpart ($S24 > 60\mu$Jy) within
the approximate 70 $\mu$m beam radius of $9\arcsec$ (which corresponds
to a relatively large chance positional coincident of about 50\% for a
24 $\mu$m source).  Of these seven, four are associated with a radio
source, two do not have a radio counterpart, and one is outside the
radio coverage.  For the three cases without a current 24 $\mu$m or
radio counterpart, the 70~$\mu$m position is located between a blend
of two or three 24 $\mu$m sources.  These blended sources could
represent valid detections at 70~$\mu$m.  All 160 $\mu$m sources have
possible 24 $\mu$m counterparts, which is not unexpected given the low
spatial resolution of the 160~$\mu$m data (where on average there are
about three 24 $\mu$m sources per 160 $\mu$m beam).  Although we find
no obvious spurious detections within the catalogs, users should be
cautious and check the images when comparing catalogs at different
wavelengths due to potential source blending.  Tables~3\&4 show the
format for an example portion of the S-COSMOS 70~$\mu$m and 160~$\mu$m
catalogs published in the on-line edition of the Journal.

PRF-fitting flux densities and aperture measurements were made using
the APEX software. The aperture and PRF measurements are in
statistical agreement.  Since PRF measurements are significantly more
accurate for faint sources \citep[e.g.,][]{fra06a}, the PRF flux
densities are used for the vast majority of sources (flag of ``p'').
Aperture measurements are adopted for sources not reasonably well
fitted by the PRF (extended sources or unresolved blends) and are
given a flag of ``a'' within the catalog.  Resolved blends fitted well
by multiple PRFs are given a flag of ``p''.  For the one blend of two
resolved sources which is not fitted well by two point sources at
70~$\mu$m, the total aperture flux is divided among the two components
based on their relative peaks (flag of ``ap'').  For consistency with
the adopted calibration of MIPS, the measurements include the aperture
correction of 1.15 for emission outside of the measured PRF
(Sec. 3.6).  No corrections for flux non-linearity have been made
\citep{sta07}.  Only one source (SCOSMOS160 J100027.0+032226, $S160
\sim 11$\,Jy) is bright enough to be significantly impacted by flux
non-linearity (the only source at 160~$\mu$m with an aperture
measurement, flag of ``a''); its flux density should be treated with
caution.

The errors on the fitted flux densities derived by APEX are not used
since they are underestimated by about a factor of 3 for these data.
We adopt errors based on the SNR measurements which represent the
fitted peak flux density divided by the uncertainty image at the
location of the source.  The errors on the measured flux densities
($S$) represent the random errors given by the SNR ratio combined with
the systematic calibration uncertainty ($\epsilon_{\rm cal}$).  The
flux density error $\sigma(S) = (1/{\rm SNR} + \epsilon_{\rm cal})S$,
where $\epsilon_{\rm cal}$ is 0.05 and 0.12 for the MIPS 70~$\mu$m
\citep{gor07} and MIPS 160~$\mu$m \citep{sta07} bands respectively.
The few sources (9) requiring large aperture measurements have
additional measurement errors of 10--20\%.

We find that the positional fitting errors calculated by APEX are 2.0
times larger on average than the expected radial positional errors of
$\approx 0.6({\rm FWHM}/{\rm SNR})$ given by \cite{con97} for all
SNRs.  We adopt the APEX errors, but treat them as $2\sigma$
uncertainties.  For these data the relationship of \cite{con97} is
appropriate for typical SNRs, but underestimates the errors for the
highest SNRs.  We adopt a floor on the uncertainty of $\epsilon_{\rm
pos} = 0.5$ pixel ($\simeq 0.1$ FWHM, Table~2), representing the
difficulty in deriving positions to better than a fractional pixel
regardless of the SNR.  The cataloged radial positional errors
($2\sigma$ uncertainties) are given by $(\sigma^2_x + \sigma^2_y +
\epsilon^2_{\rm pos})^{0.5}$, where $\sigma_x$ and $\sigma_y$ are the
fitting errors in the x-y plane derived by APEX.

\section{Results and Discussion}

\subsection{Source Counts}

The source counts were derived from SNR$\geq 5.0$ source lists corrected for
completeness over the region within the images with high coverage
($>100$ at 70~$\mu$m and $>20$ at 160~$\mu$m), corresponding to an
effective survey area for the 70 and 160~$\mu$m fields of 2.471 and
2.438 sq-deg, respectively.  The public catalogs presented in Sec. 4.2
include sources outside of this nominal area with lower coverage.  The
effective completeness value for each flux density bin was computed by
integrating over the completeness curves (Fig. 3\&4) as a function of
flux density across each bin with weights based on the measured slope
of the source counts.

Figures 5\&6 show the Euclidean-normalized differential source counts
($dN/dS\times S^{2.5}$) at 70~$\mu$m and 160~$\mu$m, and the results are
tabulated in Tables~6\&7, respectively.  The counts are calculated for
independent flux density bins.  The error bars represent the Poisson
errors associated with the number of sources and the uncertainties on
the completeness corrections.  The grey-region shows the range of
values implied from the most statistically accurate counts within each
flux density bin, including the additional uncertainty due to the
systematics associated with the calibration of MIPS (5\% at 70~$\mu$m
and 12\% at 160~$\mu$m).

At 70~$\mu$m the S-COSMOS counts are measured down to a level of
10\,mJy which is near the peak of the Euclidean-normalized
differential source counts \citep{fra06b}.  The faint ($\sim 10$\,mJy)
S-COSMOS counts at 70~$\mu$m (Fig.~5) agree with those measured for
GOODS-N \citep{fra06b} and the xFLS \citep{fra06a}.  The bright counts
for S-COSMOS also agree with the measurements from the SWIRE survey.
At intermediate flux densities of around 20--30\,mJy, the counts for
S-COSMOS are slightly lower than those found for the xFLS and model
predictions of \cite{lag04}.  All the counts have been placed on the
same scale by matching the calibration, color corrections, and the PRF
adopted for S-COSMOS.  The correction factors for the other data sets
are given in Table~5.  The previous results for the xFLS, SWIRE, and
GOODS-N did not include the aperture correction for the flux density
outside of the measured PRF.  The counts for SWIRE are based on the
public catalogs (2005 November, Data Release 3 [DR3] ) which cover 49
sq-deg.  The SWIRE counts at 70 or 160~$\mu$m have not previously been
published and are presented here to constrain the counts at the
brightest flux densities.  Only the high SNR ($\ga 10$) SWIRE sources
with completeness levels near one are presented in Figures 5\&6.

At 160~$\mu$m the S-COSMOS counts are measured down to a level of
60\,mJy.  Measurements of deeper counts are limited by confusion
(Sec. 5.2).  The measured counts at 160~$\mu$m agree with the counts
measured previously in the xFLS \citep{fra06a} and the counts derived
here based on the SWIRE survey.  As done for 70~$\mu$m, all counts are
placed on the current calibration scale (Table~5).  These correction
factors also include a decrease in the flux densities due to the new
PRF derived here (Sec. 3.4). The previous PRF images used by the xFLS
and SWIRE surveys are affected by the flux non-linearity for bright
sources at 160~$\mu$m.

The faintest ($S160<80$\,mJy) and brightest ($S160>500$\,mJy) source
counts are consistent with the \cite{lag04} model, but the observed
counts for all three surveys at intermediate flux densities are about
a factor of 1.5 times lower than the model implies.  The observed 160
$\mu$m counts show a steep increase in the differential counts for
decreasing flux densities ($dN/dS \propto S^{-3.5\pm0.2}$) for
$S160<150$\,mJy.  The slope for the faint (10--20\,mJy) S-COSMOS 70
$\mu$m source counts ($dN/dS \propto S^{-3.1\pm0.2}$) is roughly
consistent with the 160~$\mu$m slope and is slightly steeper than the
model predictions.  The differences between the model and the source
counts at 70~$\mu$m is not as clear as those seen at 160~$\mu$m.  The
results for 160~$\mu$m may suggest the importance of galaxies having
more cold dust than assumed in the models.  Observations with the
future {\em Herschel} telescope should provide better constraints on
the FIR SEDs of distant galaxies.

\subsection{Confusion Level of the MIPS 70 and 160 $\mu$\lowercase{m} Bands}

The S-COSMOS 70~$\mu$m data are dominated by instrument uncertainties
and are not deep enough to constrain the confusion noise.
\cite{fra06b} measured the MIPS 70~$\mu$m confusion noise based on the
much deeper 70~$\mu$m observations of GOODS-North.  Based on
\cite{fra06b} and the calibration scaling factors (Table~5), the
updated extragalactic confusion noise level for the MIPS 70~$\mu$m
band is $\sigma_c = 0.35\pm 0.15$\,mJy $(q=5)$, including the updated
systematic error on the flux calibration.

In contrast to the 70~$\mu$m data, the S-COSMOS 160~$\mu$m data
contain a significant noise component due to confusion.  We estimate
the confusion noise at 160~$\mu$m following the empirical technique
performed at 70~$\mu$m \citep{fra06b}.  A direct empirical measurement
of the confusion noise for the MIPS 160~$\mu$m band has not been
published previously.  \cite{dol04} report a confusion level at
160~$\mu$m based on the techniques of \cite{dol03} and the models of
\cite{lag04}.  They find that the predicted confusion noise from their
models is in reasonable agreement with the observations, assuming that
the instrumental noise ($\sigma_I$) follows the theoretical MIPS model
and integrates down as $\sigma_I \propto t^{-0.5}$.  However, the
early processing of MIPS-Ge data was not optimal, and an accurate
empirical measurement of the total instrumental noise (including
photon noise, detector noise, and noise associated with the data
processing) is required to measure the confusion level.

The instrument noise was estimated empirically by subtracting pairs of
data subsets with the same integration time and covering the exact
same region on the sky (which removes sources and any remaining cirrus
structure after filtering).  We fit the instrument noise measurements
as a function of integration time for combinations of deep pairs of
data sets and obtain $\sigma_I \propto t^{-0.49\pm0.03}$.  This result
is consistent with idealized data ($\sigma \propto t^{-0.5}$) and
highlights the success of the reduction methods in removing systematic
artifacts.  The extrapolation of instrument noise for half of the data
to the full data set yields $\sigma_{I} = 0.1134\pm
0.0033$\,MJy\,sr$^{-1}$, where the uncertainty represents the rms
measurement error combined in quadrature with the error associated with
the extrapolation.

Following the terminology of \cite{dol03}, the total noise
($\sigma_{T}$) represents the noise after the extraction of sources
above a limiting flux density ($S_{lim}$), and the photometric
confusion noise ($\sigma_{c}$) represents fluctuations due to sources
with flux densities below $S_{lim}$.  The confusion noise is given by
$\sigma_c= (\sigma_T^2-\sigma_I^2)^{0.5}$, which is appropriate for an
approximate Gaussian distribution of the noise after the extraction of
sources.  We iterate between source extraction at different limiting
flux densities and confusion noise measurements until we converge to a
solution with $q\equiv S_{\lim}/\sigma_c =5$.  For the $q=5$ solution,
we derive $\sigma_{T} = 0.1772\pm 0.0089$\,MJy\,sr$^{-1}$ and
$\sigma_{c} = 0.1362\pm 0.0119$\,MJy\,sr$^{-1}$, for a limiting source
flux density of $S160=50$\,mJy.

Noise measurements in surface brightness units (MJyr\,sr$^{-1}$)
depend on the pixel scale, and all measurements here are based on the
$8\arcsec$ pixel scale of the 160~$\mu$m image.  The uncertainties on
the noise measurements given in MJy\,sr$^{-1}$ also do not include the
12\% systematic calibration uncertainty.  Including the systematic
calibration uncertainty and the measured conversion between
surface-brightness noise and point-source noise (Sec. 4.1), the
point-source confusion level of the MIPS 160~$\mu$m band within the
COSMOS field is $\sigma_{c} = 10.0\pm 3.1$\,mJy.  The systematic
uncertainties contribute 71\% to the total error budget, while the
random errors contribute 29\% to the total error budget.

The measured confusion noise is not entirely due to galaxies.  Unlike
the case for the MIPS 70 band $\mu$m \citep[e.g.,][]{dol03,fra06b},
Galactic cirrus is not negligible at 160~$\mu$m.  We estimate the
level of confusion due to Galactic cirrus for the MIPS 160~$\mu$m band
using the background estimates from the {\em Spitzer} tool for
planning observations (SPOT) which is based on the interstellar medium
(ISM) maps from the Diffuse Infrared Background Experiment
\citep{sch98}.  At the effective wavelength of 155.9 $\mu$m of the
MIPS 160~$\mu$m band, the ISM background in the direction of the
COSMOS field is about 2\,MJy\,sr$^{-1}$.  Using the calculations of
\cite{jeo05}, this background level corresponds to an ISM confusion
noise of $\sigma_{c}({\rm ISM}) \simeq 3.4$\,mJy.  Hence, the
confusion noise due to unresolved galaxies is $\sigma_{c}({\rm gal})=
(\sigma_{c}^2({\rm total})-\sigma_{c}^2({\rm ISM}))^{0.5} = 9.4\pm
3.3$\,mJy.

The derived confusion limit agrees fairly well with predictions.
\cite{dol04} report a Source Density Criterion (SDC) limit of 40\,mJy
($\sigma_c(q_{SDC}=3.8) = 10.6$\,mJy) based on the model predictions
of \cite{lag04}.  For a direct comparison with the empirical
measurement, the Dole \& Lagache et al. predictions suggest
$\sigma_c(q=5) = 12.5$\,mJy.  We find an extragalactic confusion level
for the MIPS 160~$\mu$m band of $\sigma_c(q=5) = 9.4\pm 3.3$\,mJy,
which is just slightly lower than the predicted value.

\section{Concluding Remarks}

We present the 70 and 160~$\mu$m observations of the COSMOS field and
describe the products.  This is the first
extragalactic survey available to the public at 70 and 160~$\mu$m that
has been placed on the calibration scale derived from the recent MIPS
calibration papers.  We provide updated correction factors for the
previously released catalogs of the xFLS \citep{fra06a} and SWIRE
\citep{lon04} programs.  Counts are presented based on the S-COSMOS
and previous surveys and are found to be in reasonable agreement with
the model of \cite{lag04}.  However, the faint 160~$\mu$m source
counts are significantly steeper than model predictions.  We measure
an empirical extragalactic confusion noise level of $\sigma_c =
9.4\pm3.3$\,mJy ($q=5$) for the MIPS 160~$\mu$m band.  In comparison,
the expected confusion noise at 160~$\mu$m for {\em Herschel} is about
1.5--2\,mJy ({\em Herschel} Observation Planning Tool, HSPOT, version
3.4).  Future observations with the {\em Herschel} telescope should
constrain the counts and far-infrared properties better than can be
done currently due to confusion for the MIPS 160~$\mu$m band.

\acknowledgments

We thank our colleagues associated with the {\em Spitzer} mission who
have made these observations possible.  This work is based on
observations made with the {\em Spitzer Space Telescope}, which is
operated by the Jet Propulsion Laboratory, California Institute of
Technology under a contract with NASA. Support for this work was
provided by NASA through an award issued by JPL/Caltech.

\clearpage
\begin{deluxetable}{lrcr}
\tablewidth{0pt}
\tablecaption{{\em Spitzer} COSMOS Observations with MIPS}
\tablehead{
\colhead{Epoch}&\colhead{Dates}&\colhead{Campaign}&\colhead{AORs}}
\startdata

Cycle-2   &  2006 Jan 06--09&   MIPS006300  &   21\\
Cycle-3a  &  2007 Jan 01--13&   MIPS010800  &   70\\
Cycle-3b  &  2007 May 18--28&   MIPS011900  &   61\\
Cycle-3c  &  2008 Jan 06--07&   MIPS013500  &    5\\

\enddata
\tablecomments{{\em Spitzer} programs 20070 and 30143.}
\end{deluxetable}

\begin{deluxetable}{lcc}
\tablewidth{0pt}
\tablecaption{Summary of Product Properties}
\tablehead{
\colhead{Property}&\colhead{70~$\mu$m}&\colhead{160~$\mu$m}}
\startdata

Image pixel scale & $4\arcsec$ & $8\arcsec$ \\
Image FWHM resolution & $18.6\arcsec$ & $39\arcsec$\\
Nominal coverage level& 100 & 20 \\
Median coverage$^{a}$ & 140.3  & 28.3\\
Median exposure time$^{a}$ & 1350\,s  & 273\,s\\
Surface-brightness noise ($1\sigma$)$^{a}$ & 0.13 MJy\,sr$^{-1}$ & 0.18 MJy\,sr$^{-1}$\\
Point-source noise ($1\sigma$)$^{a}$ & 1.7\,mJy &  13\,mJy \\
Area mapped$^{a}$ & 2.471 sq-deg  & 2.438 sq-deg \\
Number of sources ($\geq 5.0\sigma$)& 1512 & 499 \\
\enddata

\tablecomments{$^{a}$Measured properties for coverages (number of
  observations) larger than the nominal coverage level.}
\end{deluxetable}

\begin{deluxetable}{lrrrrrrc}
\tabletypesize{\footnotesize} 
\tablewidth{0pt}
\tablecaption{{\em Spitzer} COSMOS 70~$\mu\lowercase{m}$ Catalog}
\tablehead{
\colhead{Source Name}&
\colhead{$\alpha$(J2000)} &
\colhead{$\delta$(J2000)} &
\colhead{Error} &
\colhead{$S70$}&
\colhead{$\sigma(S70)$}&
\colhead{SNR}&
\colhead{Flag}\\
&
\colhead{(deg)}&
\colhead{(deg)}&
\colhead{(arcsec)}&
\colhead{(mJy)}&
\colhead{(mJy)}&
\colhead{}&
\colhead{ }\\
\colhead{(1)}&
\colhead{(2)}&
\colhead{(3)}&
\colhead{(4)}&
\colhead{(5)}&
\colhead{(6)}&
\colhead{(7)}&
\colhead{(8)}
}

\startdata

SCOSMOS70 J095546.9+013605&  148.945497&  1.601560&  5.0&  20.6&   4.9 &  5.3 &  p\\
SCOSMOS70 J095552.5+014243&  148.968967&  1.712061&  2.2& 226.2&  21.9 & 21.3 &  p\\
SCOSMOS70 J095605.9+014443&  149.024976&  1.745507&  4.7&  17.1&   3.9 &  5.7 &  p\\
SCOSMOS70 J095610.1+014213&  149.042376&  1.703652&  4.8&  14.3&   3.4 &  5.4 &  p\\
SCOSMOS70 J095615.0+014315&  149.062575&  1.720973&  4.0&  16.6&   3.2 &  7.1 &  p\\

\enddata

\tablecomments{The entire S-COSMOS 70~$\mu$m catalog ($\geq
  5.0\sigma$) is presented in the electronic edition of the
  Astronomical Journal. A portion of Table~3 is shown here for
  guidance regarding the form and content of the catalog.  Column (1)
  gives the source names following the IAU designations.  Column (2)
  and column (3) are the right ascension ($\alpha$) and declination
  ($\delta$) J2000.0 source positions in decimal degrees.  Column (4)
  is the radial positional uncertainty in arcsec ($2\sigma$).  Column
  (5) is the total flux density measurement in mJy.  Column (6) is the
  flux density error in mJy ($1\sigma$), including the systematic
  uncertainty of the absolute flux density scale.  Column (7) is the
  SNR of the peak.  Column (8) is the flag for the flux density
  measurement method.  A flag of ``p'' indicates that the source was
  fitted by the PRF, ``a'' indicates an aperture measurement with a
  diameter of $96\arcsec$, and ``ap'' indicates an aperture
  measurement whose flux was divided between two components based on
  the relative strengths of their peaks.}

\end{deluxetable}

\begin{deluxetable}{lrrrrrrc}
\tabletypesize{\footnotesize} 
\tablewidth{0pt}
\tablecaption{{\em Spitzer} COSMOS 160~$\mu\lowercase{m}$ Catalog}
\tablehead{
\colhead{Source Name}&
\colhead{$\alpha$(J2000)} &
\colhead{$\delta$(J2000)} &
\colhead{Error} &
\colhead{$S160$}&
\colhead{$\sigma(S160)$}&
\colhead{SNR}&
\colhead{Flag}\\
&\colhead{(deg)}&
\colhead{(deg)}&
\colhead{(arcsec)}&
\colhead{(mJy)}&
\colhead{(mJy)}&
\colhead{}&
\colhead{ }\\
\colhead{(1)}&
\colhead{(2)}&
\colhead{(3)}&
\colhead{(4)}&
\colhead{(5)}&
\colhead{(6)}&
\colhead{(7)}&
\colhead{(8)}
}
\startdata

SCOSMOS160 J095619.9+012425&  149.083255&  1.407074&  9.2&   134.9&    40.1&   5.6&   p\\
SCOSMOS160 J095621.2+015951&  149.088411&  1.997640&  9.7&   122.5&    37.2&   5.4&   p\\
SCOSMOS160 J095635.5+014332&  149.147947&  1.725679&  9.2&   122.7&    35.4&   5.9&   p\\
SCOSMOS160 J095635.7+012544&  149.149042&  1.429048&  4.7&  1073.7&   188.5&  18.0&   p\\
SCOSMOS160 J095636.9+015415&  149.153962&  1.903362&  5.0&  1093.4&   209.8&  13.9&   p\\

\enddata

\tablecomments{The entire S-COSMOS 160~$\mu$m catalog ($\geq
  5.0\sigma$) is presented in the electronic edition of the
  Astronomical Journal.  A portion of Table~4 is shown here for
  guidance regarding the form and content of the catalog.  Column
  headers are the same as described in Table~3 for columns
  (1)--(7). Column (8) is the flag for the flux density measurement
  method where ``p'' indicates a PRF measurement and ``a'' indicates
  an aperture measurement with a diameter of $4\arcmin$.}

\end{deluxetable}

\begin{deluxetable}{lrcr}
\tablewidth{0pt}
\tablecaption{Scaling Factors for MIPS-Ge Surveys}
\tablehead{
\colhead{Survey, Version}&\colhead{70~$\mu$m}&\colhead{160~$\mu$m}}
\startdata

S-COSMOS, version 1\&2& 1.25  & 1.07\\
S-COSMOS, version 3   & 1.00  & 1.00 \\
xFLS\citep{fra06a}    & 1.20  & 0.97 \\
SWIRE (2005 November, DR3) & 1.10  & 0.98 \\
GOODS-N\citep{fra06b} & 1.15  & \dotfill \\

\enddata 

\tablecomments{Multiplicative scaling factors required to place
catalogs on the same calibration scale derived for the official
calibration of MIPS \citep{gor07,sta07} and used here for the
S-COSMOS, version 3 products.  Factors include the combination of
color-corrections, calibration updates, PRF updates, and the aperture
correction for emission outside of the measured PRF. }

\end{deluxetable}

\begin{deluxetable}{ccccccc}
\tablewidth{0pt}
\tablecaption{S-COSMOS 70$\mu\lowercase{m}$ Source Counts}
\tablehead{
\colhead{Observed}&
\colhead{}&
\colhead{}&
\colhead{Observed}&
\colhead{}&
\colhead{Eddington}&
\colhead{}\\
\colhead{Average S$_{\nu}$} &
\colhead{S$_{\rm low}$}&
\colhead{S$_{\rm high}$} &
\colhead{Number} &
\colhead{Completeness} &
\colhead{Bias}&
\colhead{$dN/dS\,S^{2.5}$}\\
\colhead{(mJy)}&
\colhead{(mJy)}&
\colhead{(mJy)}&
\colhead{}&
\colhead{}&
\colhead{}&
\colhead{(gal sr$^{-1}$ Jy$^{1.5}$)}
}
\startdata

   11.02 &    10 &    12 &   245 &  0.68$\pm$  0.05 & 1.06&   2872$\pm$ 280\\
   12.84 &    12 &    14 &   186 &  0.90$\pm$  0.03 & 1.04&   2478$\pm$ 199\\
   15.30 &    14 &    17 &   180 &  0.98$\pm$  0.02 & 1.03&   2304$\pm$ 178\\
   19.08 &    17 &    22 &   144 &  1.00$\pm$  0.02 & 1.01&   1937$\pm$ 166\\
   25.76 &    22 &    30 &   108 &  1.00$\pm$  0.02 & 1.01&   1924$\pm$ 189\\
   36.26 &    30 &    45 &    92 &  1.00$\pm$  0.02 & 1.00&   2086$\pm$ 221\\
   54.05 &    45 &    70 &    46 &  1.00$\pm$  0.02 & 1.00&   1697$\pm$ 252\\
   86.36 &    70 &   110 &    22 &  1.00$\pm$  0.02 & 1.00&   1637$\pm$ 350\\

\enddata

\tablecomments{The counts are based on the $\geq 5.0\sigma$ catalog
  and were measured for an area of 2.471\,sq-deg with a coverage of
  greater than 100 observations. The counts were corrected for
  completeness and the Eddington bias ($S_{\rm obs}/S_{\rm true}$).
  The tabulated uncertainties for the differential counts
  ($dN/dS\,S^{2.5}$) include the Poisson noise and the completeness
  uncertainties, but do not include the additional systematic
  uncertainty (5\%) associated with the calibration of the MIPS
  70~$\mu$m band.}

\end{deluxetable}

\begin{deluxetable}{ccccccc}
\tablewidth{0pt}
\tablecaption{S-COSMOS 160~$\mu\lowercase{m}$ Source Counts}
\tablehead{
\colhead{Observed}&
\colhead{}&
\colhead{}&
\colhead{Observed}&
\colhead{}&
\colhead{Eddington}&
\colhead{}\\
\colhead{Average S$_{\nu}$} &
\colhead{S$_{\rm low}$}&
\colhead{S$_{\rm high}$} &
\colhead{Number} &
\colhead{Completeness} &
\colhead{Bias}&
\colhead{$dN/dS\,S^{2.5}$}\\
\colhead{(mJy)}&
\colhead{(mJy)}&
\colhead{(mJy)}&
\colhead{}&
\colhead{}&
\colhead{}&
\colhead{(gal sr$^{-1}$ Jy$^{1.5}$)}
}
\startdata

   65.05 &    60 &    70 &    82 &  0.63$\pm$  0.05 & 1.09 & 16619$\pm$2260\\
   74.82 &    70 &    80 &    83 &  0.85$\pm$  0.03 & 1.08 &  17875$\pm$2060\\
   89.01 &    80 &   100 &    83 &  0.95$\pm$  0.02 & 1.06 &  12713$\pm$1420\\
  139.18 &   120 &   160 &    40 &  1.00$\pm$  0.02 & 1.02 &   9445$\pm$1505\\
  187.72 &   160 &   250 &    29 &  1.00$\pm$  0.02 & 1.01 &   6526$\pm$1218\\

\enddata

\tablecomments{The counts are based on the $\geq 5.0\sigma$ catalog
  and were measured for an area of 2.438\,sq-deg with a coverage of
  greater than 20 observations. The counts were corrected for
  completeness and the Eddington bias ($S_{\rm obs}/S_{\rm
  true}$). The uncertainties for the differential counts
  ($dN/dS\,S^{2.5}$) include the Poisson noise and the completeness
  uncertainties, but do not include the additional systematic
  uncertainty (12\%) associated with the calibration of the MIPS
  160~$\mu$m band. }

\end{deluxetable}

%use below for emulateapj preprint
%\epsscale{0.70}

%\clearpage
%\begin{figure}
%\plotone{f1.ps}
%\vspace*{1.5cm}
\begin{figure}[!hbt]
\centering
\includegraphics[width=0.9\textwidth]{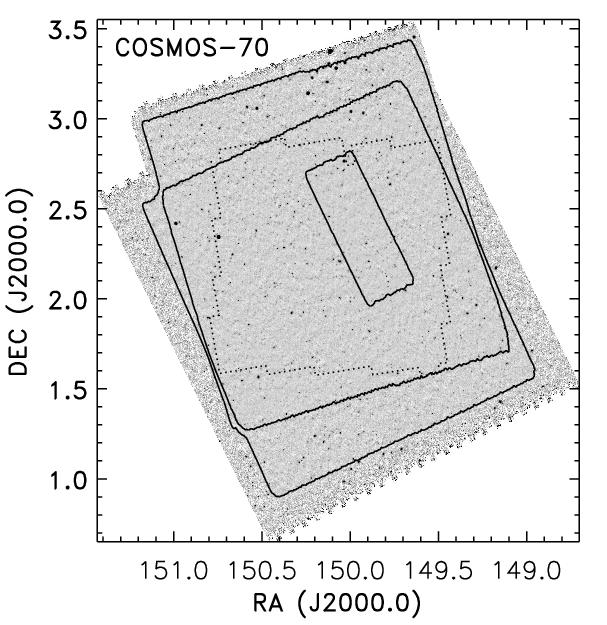}
\caption{The S-COSMOS 70~$\mu$m image shown on a logarithmic
  grey-scale.  The solid lines show contours representing coverages of
  50, 100, and 250 observations.  The central rectangle (highest
  coverage $\geq 250$) shows the Cycle-2 test field, and the dotted
  line shows the extent of the {\em HST} ACS field.}
\end{figure}

%\clearpage
%\begin{figure}
%\plotone{f2.ps}
%\vspace*{1.5cm}
\begin{figure}[!hbt]
\centering
\includegraphics[width=0.9\textwidth]{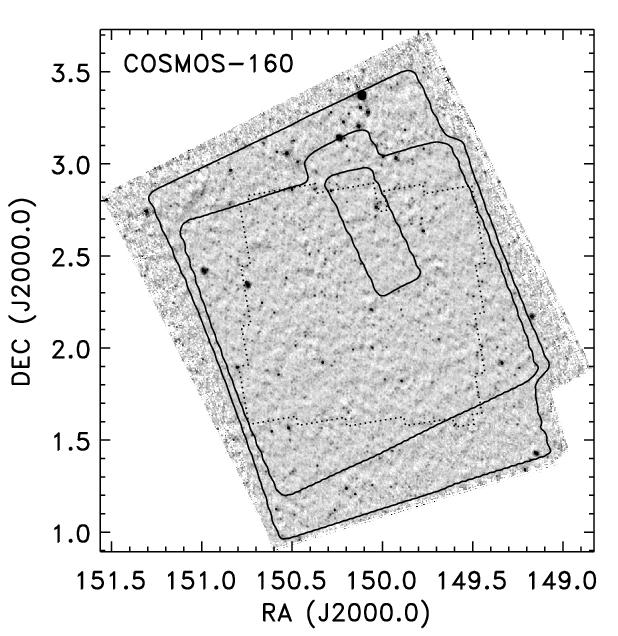}
\caption{The S-COSMOS 160~$\mu$m image shown on a logarithmic
  grey-scale.  The solid lines show contours representing a coverage
  of 10, 20, and 50 observations.  The deep central contour shows the
  Cycle-2 test field, and the dotted line shows the extent of the {\em
  HST} ACS field.}
\end{figure}

%\epsscale{0.90}

%\clearpage
%\begin{figure}
%\plotone{f3.ps}

\begin{figure}
\centering
\includegraphics[width=0.9\textwidth]{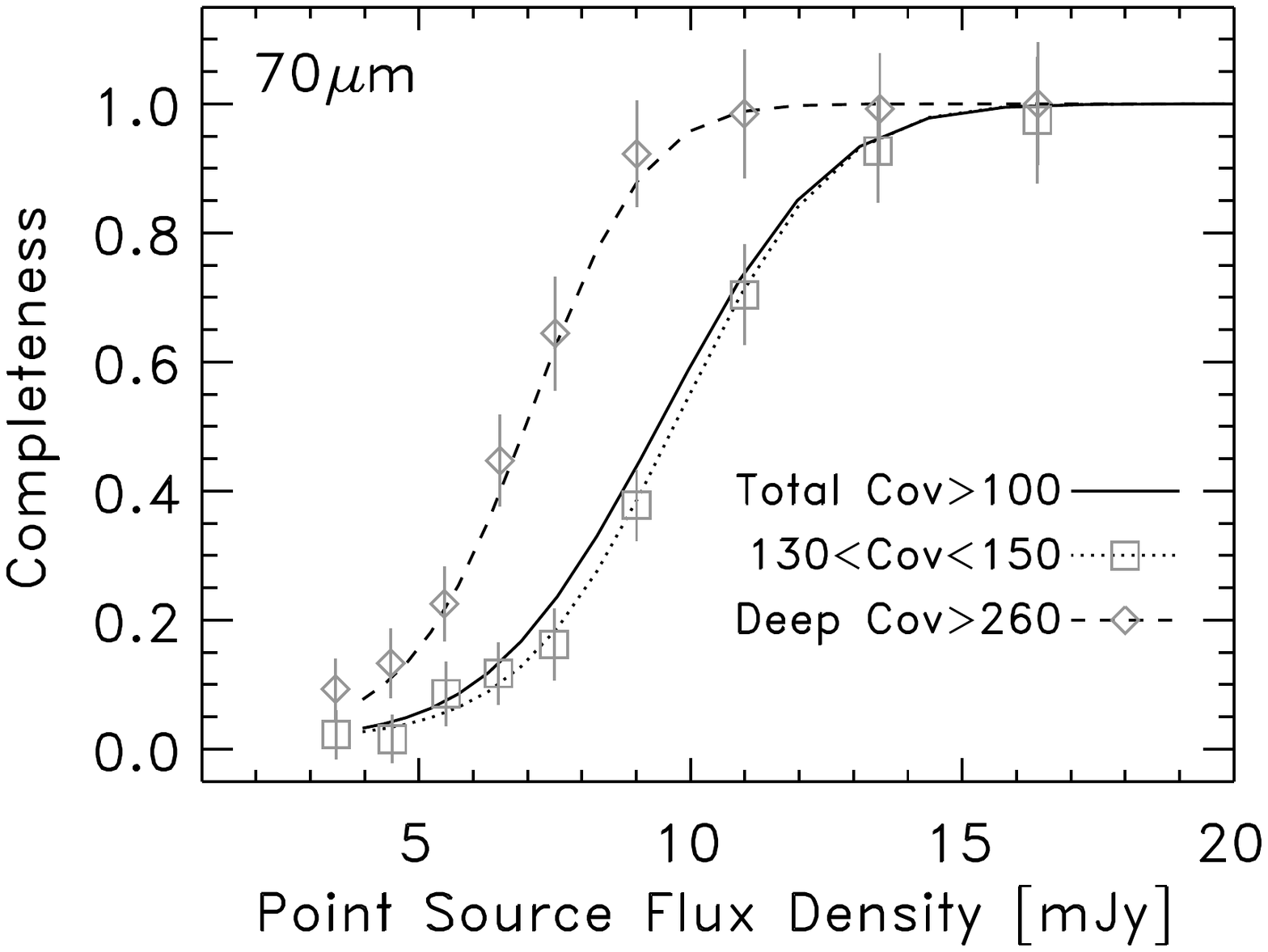}
\vspace*{-2cm}
\caption{The 70~$\mu$m completeness levels (SNR $\geq 5.0$) as a function of
  flux density and coverage.  The effective average completeness level
  used for the source counts is given by the solid line (Total
  Cov$>100$).  Simulation results for the deep area (Cov $>260$) and
  typical coverage values ($130<{\rm Cov}<150$) are shown by the grey
  diamonds and squares.  The dashed and dotted lines represent the
  expected completeness level based on the SNR threshold for the
  regions used in the simulations.}
\end{figure}

%\clearpage
%\begin{figure}
%\plotone{f4.ps}
%\vspace*{1.5cm}
\begin{figure}
\centering
\includegraphics[width=0.9\textwidth]{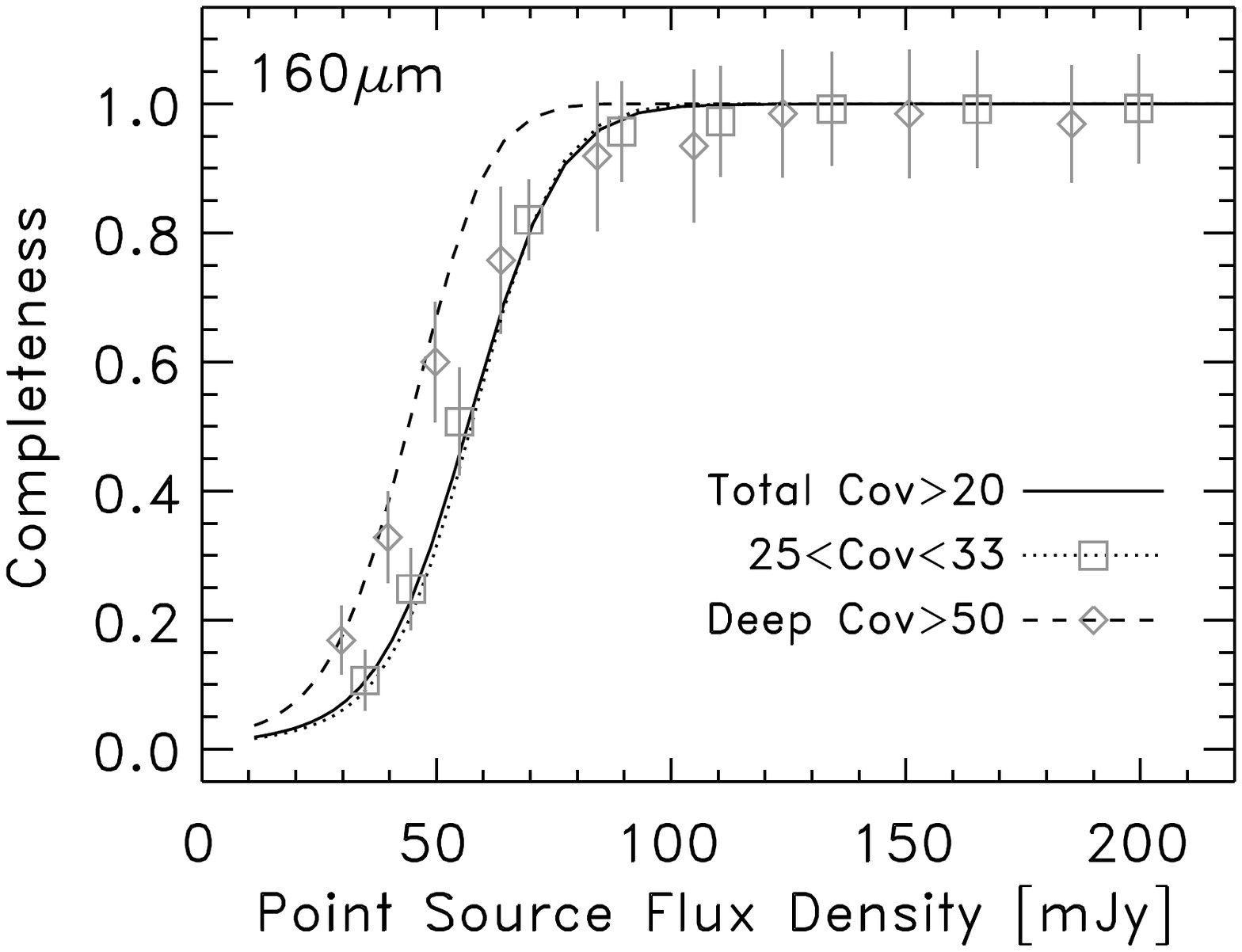}
\vspace*{-2cm}
\caption{The 160~$\mu$m completeness levels (SNR $\geq 5.0$) as a
  function of flux density and coverage.  The effective average
  completeness level used for the source counts is given by the solid
  line (Total Cov $>20$).  Simulation results for the deep area
  (Cov $>50$) and typical coverage values ($25<{\rm Cov}<33$) are shown
  by the grey diamonds and squares, along with the expected
  completeness level based on the SNR threshold and coverage values
  (dashed and dotted lines respectively)}.

\end{figure}

%\clearpage
%\begin{figure}
%\plotone{f5.ps}
%\vspace*{1.5cm}
\begin{figure}
\centering
\includegraphics[width=0.9\textwidth]{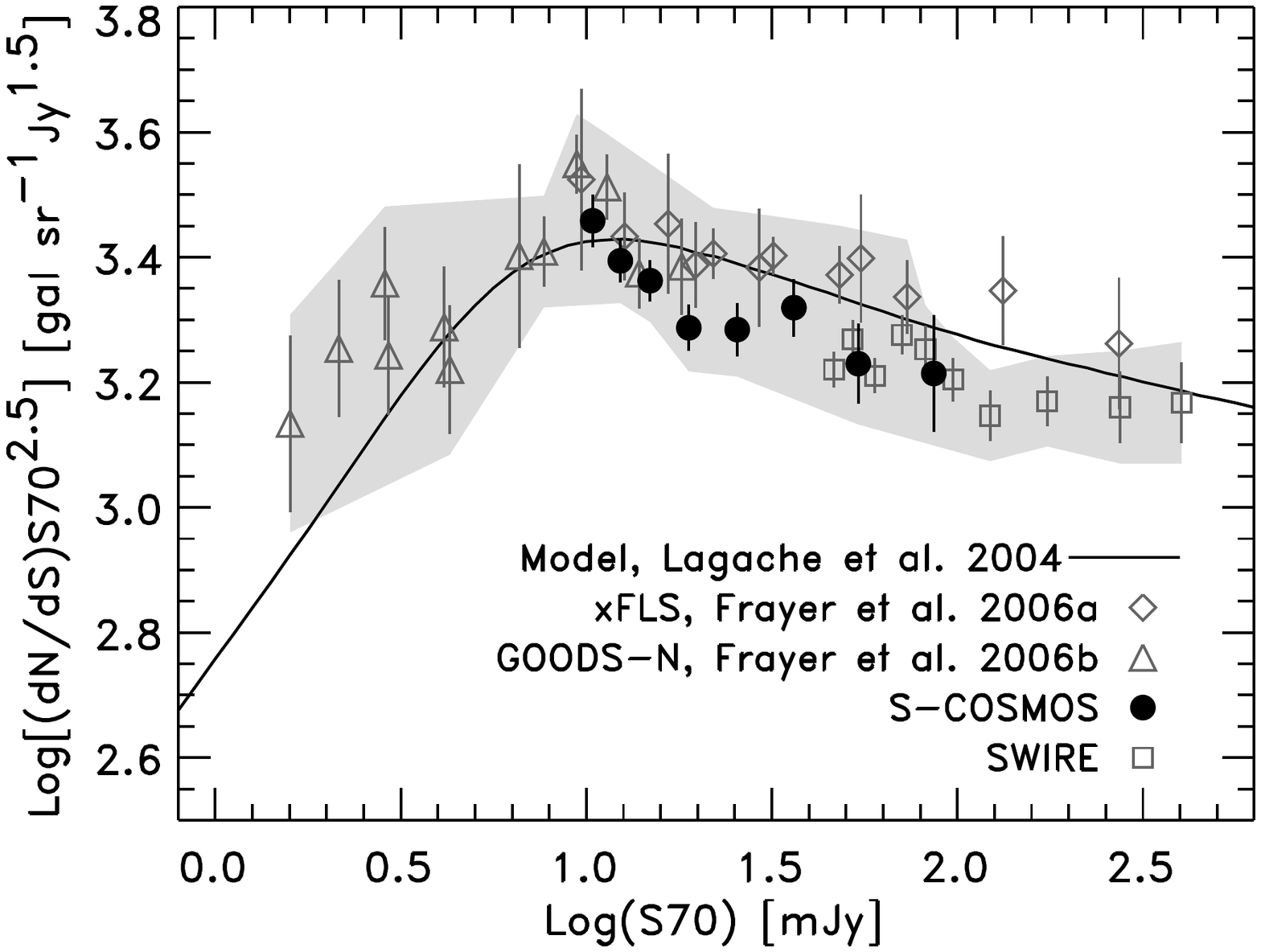}
\vspace*{-2cm}
\caption{The Euclidean-normalized differential source counts at
  70~$\mu$m compared to the evolutionary model of \cite{lag04} (solid
  line).  The S-COSMOS and SWIRE counts are new measurements.  The
  xFLS and GOODS-N counts are from \cite{fra06a,fra06b} respectively.
  All data are on the same calibration scale.  The grey region shows
  the range of values for the accurate data points
  within each flux bin and includes the additional systematic uncertainty
  associated with the calibration at 70~$\mu$m (5\%).}
\end{figure}

%\clearpage
%\begin{figure}
%\plotone{f6.ps}
%\vspace*{1.5cm}
\begin{figure}
\centering
\includegraphics[width=0.9\textwidth]{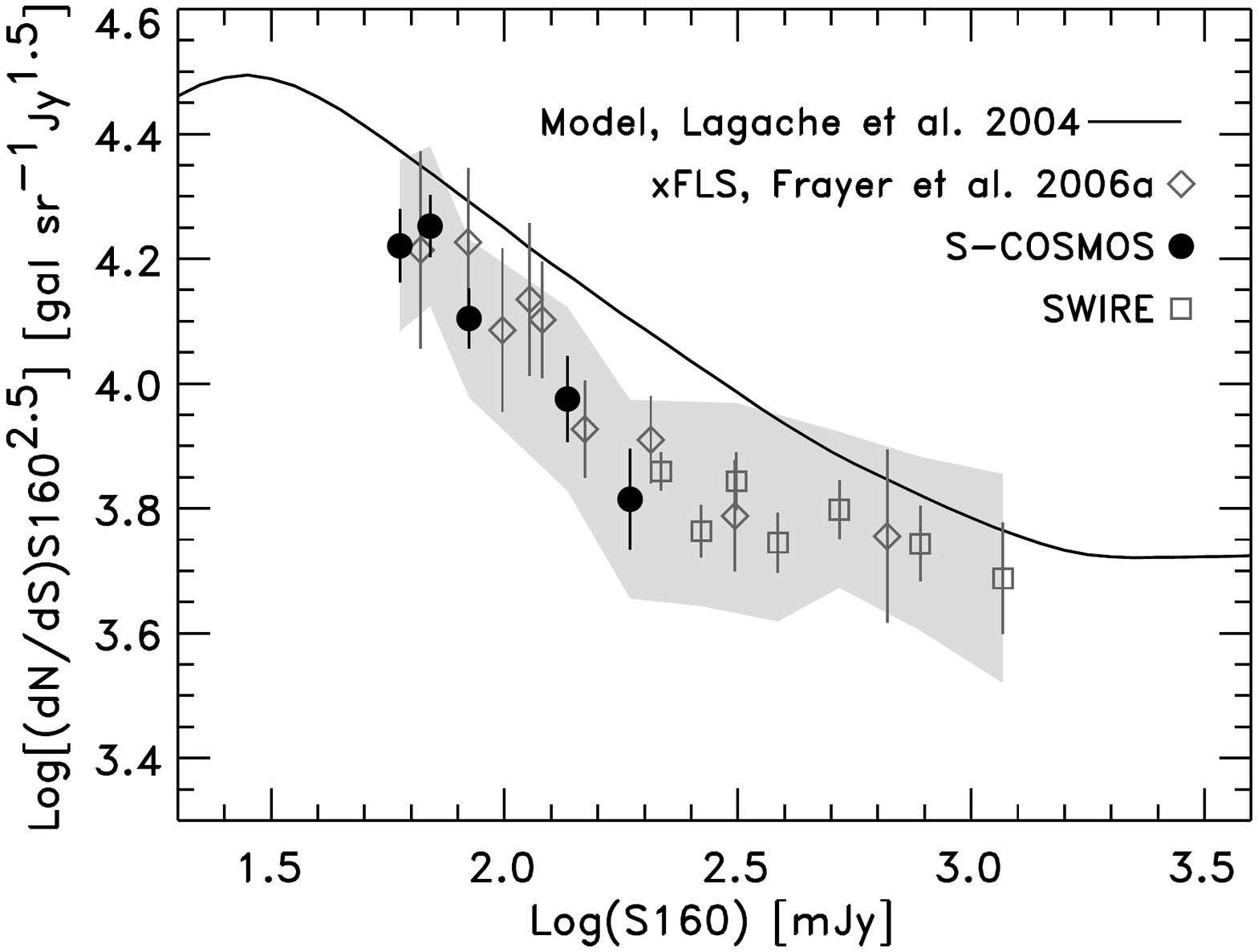}
\vspace*{-2cm}
\caption{The Euclidean-normalized differential source counts at
  160~$\mu$m compared to the evolutionary model of \cite{lag04} (solid
  line).  The S-COSMOS and SWIRE counts are new measurements, while
  the xFLS counts are from \cite{fra06a}.  All data are on the same
  calibration scale.  The grey region shows the range of values for
  the accurate data points within each flux bin and includes
  the additional systematic uncertainty associated with the
  calibration at 160~$\mu$m (12\%).}
\end{figure}


\begin{thebibliography}{}


\bibitem[Capak et al.(2007)]{cap07} Capak, P., et al.\ 2007, \apjs,
172, 99

\bibitem[Condon(1997)]{con97} Condon, J.~J.\ 1997, \pasp, 
109, 166 


\bibitem[Dale et al.(2005)]{dal05} Dale, D.~A., et al.\ 2005, 
\apj, 633, 857 

\bibitem[Dole et al.(2004)]{dol04} Dole, H., et al.\ 2004, 
\apjs, 154, 93 

\bibitem[Dole et al.(2003)]{dol03} Dole, H., Lagache, G., 
\& Puget, J.-L.\ 2003, \apj, 585, 617 

\bibitem[Engelbracht et al.(2007)]{eng07} Engelbracht, C.~W., et al.\
2007, \pasp, 119, 994

\bibitem[Frayer et al.(2006a)]{fra06a} Frayer, D.~T., et al.\ 
2006a, \aj, 131, 250 

\bibitem[Frayer et al.(2006b)]{fra06b} Frayer, D.~T., et al.\ 
2006b, \apjl, 647, L9 

\bibitem[Gordon et al.(2007)]{gor07} Gordon, K.~D., et al.\ 
2007, \pasp, 119, 1019 

\bibitem[Gordon et al.(2005)]{gor05} Gordon, K.~D., et al.\ 
2005, \pasp, 117, 503 

\bibitem[Hasinger et al.(2007)]{has07} Hasinger, G., et al.\ 2007,
\apjs, 172, 29

\bibitem[Jeong et al.(2005)]{jeo05} Jeong, W.-S., Mok Lee, 
H., Pak, S., Nakagawa, T., Minn Kwon, S., Pearson, C.~P., 
\& White, G.~J.\ 2005, \mnras, 357, 535 

\bibitem[Klaas et al.(2001)]{kla01} Klaas, U., et al.\ 2001, \aap,
379, 823

\bibitem[Koekemoer et al.(2007)]{koe07} Koekemoer, A.~M., et al.\
2007, \apjs, 172, 196

\bibitem[Krist(2002)]{kri02} Krist, J. 2002, Tiny Tim/{\em SIRTF} User's Guide
(Pasadena: SSC)

\bibitem[Lagache et al.(2004)]{lag04} Lagache, G., et al.\ 
2004, \apjs, 154, 112 

\bibitem[Lonsdale et al.(2004)]{lon04} Lonsdale, C., et al.\ 
2004, \apjs, 154, 54 


\bibitem[Makovoz(2004)]{mak04} Makovoz, D.\ 2004, \pasp, 116, 971

\bibitem[Makovoz \& Marleau(2005)]{mak05} Makovoz, D., \& Marleau,
F.~R.\ 2005, \pasp, 117, 1113

\bibitem[Sanders et al.(2007)]{san07} Sanders, D.~B., et al.\ 
2007, \apjs, 172, 86 

\bibitem[Schinnerer et al.(2007)]{sch07} Schinnerer, E., et al.\ 2007,
\apjs, 172, 46

\bibitem[Schlegel et al.(1998)]{sch98} Schlegel, D.~J., 
Finkbeiner, D.~P., \& Davis, M.\ 1998, \apj, 500, 525

\bibitem[Scoville et al.(2007)]{sco07} Scoville, N., et al.\ 2007,
\apjs, 172, 38

\bibitem[Stansberry et al.(2007)]{sta07} Stansberry, J.~A., 
et al.\ 2007, \pasp, 119, 1038 

\end{thebibliography}
\end{document}